\documentclass[11pt]{article}
\usepackage{moriond,epsfig}

\def\Journal#1#2#3#4{{#1} {\bf #2}, #3 (#4)}

\def\PLB{{\em Phys. Lett.}  B}
\def\PRL{\em Phys. Rev. Lett.}
\def\PRD{{\em Phys. Rev.} D}

\def\be{\begin{equation}}
\def\ee{\end{equation}}
\def\bea{\begin{eqnarray}}
\def\eea{\end{eqnarray}}

\begin{document}
\vspace*{4cm}
\title{NON SUSY SEARCHES AT THE TEVATRON}

\author{G.S. Muanza\\
on behalf of the D0 and CDF Collaborations\vspace*{0.5cm}}

\address{IPN Lyon, CNRS-IN2P3, Universit\'e de Lyon\\
4, Rue E. Fermi - B\^at. 210 - 69622 Villeurbanne Cedex - France}

\maketitle\abstracts{
Recent searches for non-SUSY exotics in $p\bar{p}$ collisions at a center-of-mass energy
of $1.96$ TeV at the Tevatron Run II are reported. The emphasis is put on the results
of model-driven analyses which were updated to the full Run IIA datasets corresponding
to integrated luminosities of about 1 $fb^{-1}$.}

\section{Introduction}
\par
Numerous searches for non Higgs and SUSY extensions of the Standard Model (SM) are conducted at the
Tevatron Run II.
\par\noindent
In this report we concentrate on the model-driven analyses that were recently updated by the D0
and CDF collaborations to the full Run IIA datasets, representing integrated luminosities
slightly in excess of 1 $fb^{-1}$. Details about these analyses can be found in reference \cite{TEVATRON-NP}.
\par
The SM is constructed with the following ingredients: it's a quantum field theory
where the matter fields are replicated into three families of quarks and leptons. This field theory is
placed into a four-dimensional space-time. Its lagrangian is invariant under the Poincar\'e group and
under the $SU_{C}(3)\times SU_{L}(2)\times U_{Y}(1)$ gauge group. The electroweak symmetry
breaking is provided by the Higgs mechanism.
\par
For about three decades all the experimental tests of the SM have shown no significant deviations with 
respect to its predictions. However the SM leaves many questions unresolved and is clearly not a full and
a satisfactory theory. Therefore many ideas have been proposed to try to extend both 
predictivity and its domain of validity. 
\par\noindent
Among these ideas is a possible sub-structure of the particles considered as elementary in the SM. This could
explain the replication of the quarks and leptons into three families.
\par\noindent
Another path is a possible extension
of the SM gauge symmetries. This enables to envisage a unification of the three fundamental
interactions described by the SM at a very high energy scale, whilst explaining their differences at low
energy as results of different symmetry breakings.
\par\noindent
One can also postulate the existence of extra space dimensions that
could explain the hierarchy between the Planck and the electroweak scales as well as the relative
weakness of the gravitational interaction with respect to the three other fundamental interactions.
\par
Searches for fermions sub-structure are reported in section ~\ref{sec:fermion_substruct},
section ~\ref{sec:extd_gauge_sym} and section ~\ref{sec:extd_space_dim} contain searches for 
hints of extended gauge symmetries and of extra space dimensions respectively.
\par\noindent
All the exclusion limits are given at the $95\%$ confidence level.
%
%%%
%
\section{Fermions Sub-Structure}\label{sec:fermion_substruct}
\par
In this section, we describe searches driven by two types of models that can be related
to a fermions sub-structure. Searches for leptoquarks are presented in the first sub-section
and the second sub-section contains searches for fermion compositeness.
\subsection{Search for Leptoquarks}
The leptoquarks ($LQ$) carry both a lepton and a baryon quantum number. The relevant phenomenological
parameters are $M_{LQ}$ mass for the scalar leptoquarks (simply denoted $LQ$) and 
in addition two anomalous couplings for the vector leptoquarks (denoted $VLQ$).
\par
\begin{figure}[h]
\begin{center}
\psfig{figure=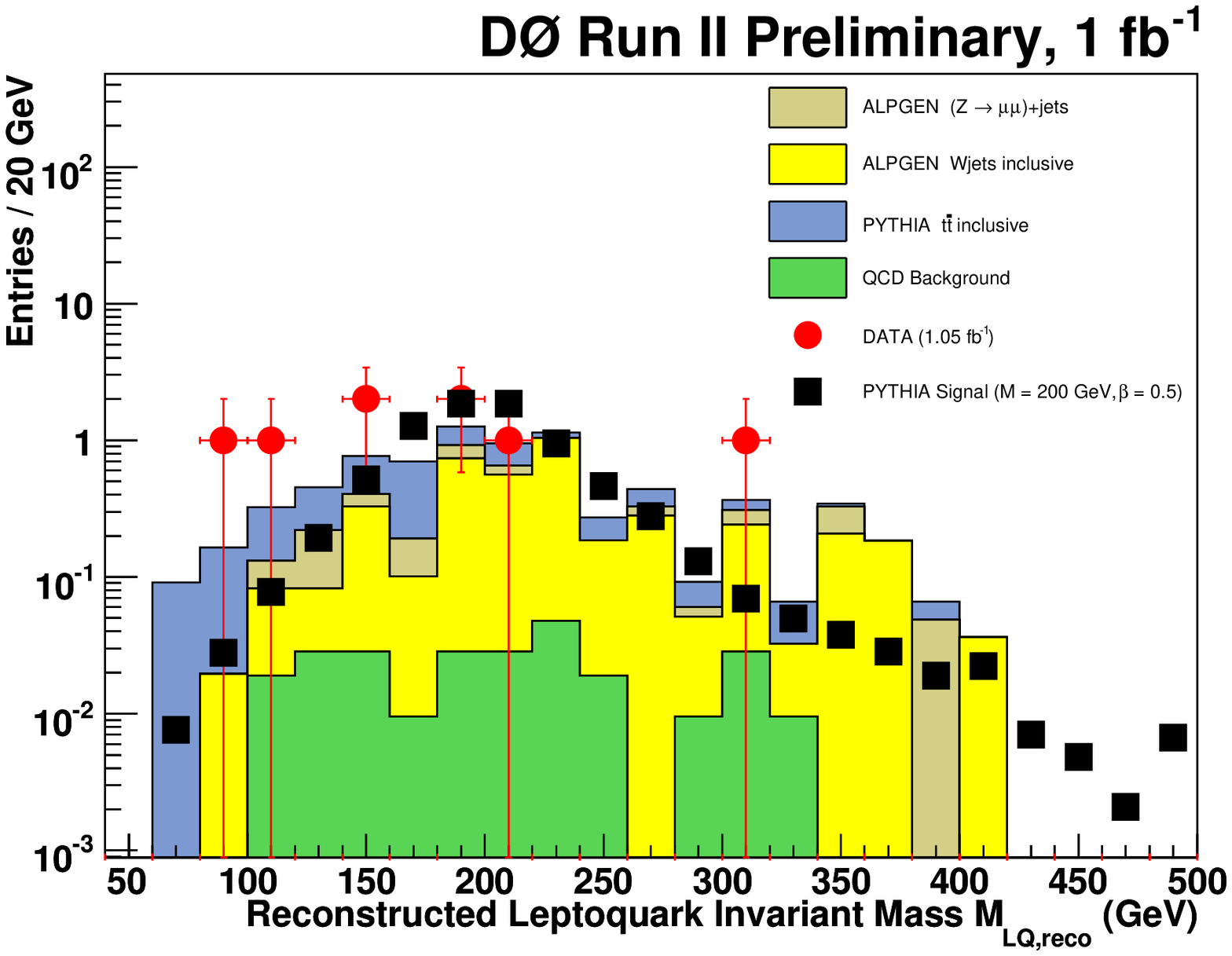,height=4cm}
\hspace{0.4cm}
\psfig{figure=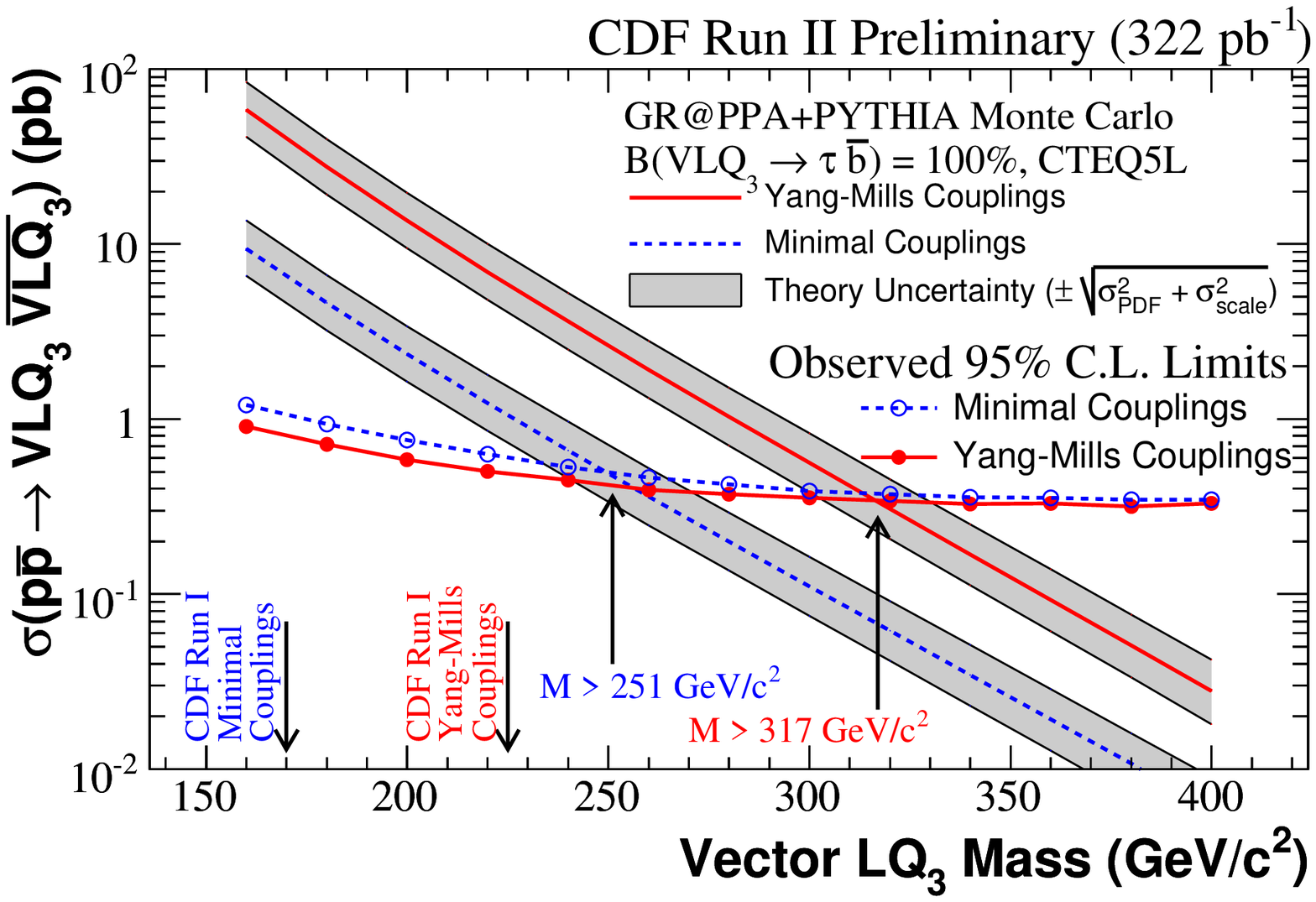,height=4cm}
\caption{Distribution of the $M_{LQ_{2}}$ in a $LQ_{2}+\bar{LQ_{2}}\to\mu\nu+q'\bar q$ search
performed by D0 (left). Limit on the production cross section of
$VLQ_{3}+\bar{VLQ_{3}}\to\tau^{+}\tau^{-}b\bar{b}$ established
by CDF (right).\label{fig:LQ}}
\end{center}
\end{figure}

D0 performed a search for second
generation scalar $LQ$ in the $LQ_{2}+\bar{LQ_{2}}\to\mu\nu+q'\bar q$ channel,
in a dataset of $\int{\cal L}dt=1.05 fb^{-1}$. Events
containing a hard and isolated muon plus jets and large missing transverse energy ($\rlap{\kern0.25em/}E_{T}$) and 
$H_{T}$ are selected. The left hand side of figure~\ref{fig:LQ} shows the $LQ_{2}$ reconstructed mass where the data
are background-like. A limit excluding $M_{LQ_{2}}<210$ GeV is set in the hypothesis of this
semi-leptonic decay of the $LQ_{2}$ pairs.
\par
CDF analysed 322 $pb^{-1}$ to search for third generation $VLQ$ in the channel
$VLQ_{3}+\bar{VLQ_{3}}\to\tau^{+}\tau^{-}b\bar{b}$ whith one $\tau$ subsequently decaying
into hadrons and the other into leptons. An $H_{T}$ variable summing up the
$\rlap{\kern0.25em/}E_{T}$ and the $p_{T}$ of each reconstructed object in the
studied topology is used to discriminate the signal from the background. Its distribution does 
not reveal any data excess. This enables to set limits excluding $M_{VLQ_{3}}<251$ GeV and 
$M_{VLQ_{3}}<317$ GeV for minimal and Yang-Mills couplings respectively as displayed
in the right hand part of figure~\ref{fig:LQ}.

%
%%%
%
\subsection{Search for Leptons or Quarks Compositeness}\label{sec:composite}
\par
For the hypothesis of quarks and leptons compositeness as detailed
in reference \cite{baur-comp}, the main parameters are the excited fermion mass $M_{f^{*}}$
and the compositeness scale $\Lambda$.
\par
We report two searches for excited electrons and muons carried by D0 with integrated luminosities
of $1 fb^{-1}$ and $0.38 fb^{-1}$ respectively. These excited leptons ($\ell^{*\pm}$) are produced by 
the following contact interaction process: $q\bar q\to \ell^{*\pm}\ell^{\mp}$ and subsequently decay into
the $\ell^{*\pm}\to\gamma +\ell^{\pm}$ mode, leading to $\gamma+\ell^{\pm}\ell^{\mp}$ final states.
The analyses essentially consist in selecting events with two hard and isolated leptons plus a hard
and isolated photon and to search for a resonance in the $M_{\gamma\ell^{\pm}}$ distribution.  
\par\noindent
Since no excess of data is found with respect to the SM background, exclusion limits 
are derived in the $\Lambda$ and $M_{\ell^{*}}$ plane. 
For example, for $\Lambda=1$ TeV: $M_{e^{*}}>756$ GeV and $M_{\mu^{*}}>618$ GeV.
\par
We also report that in the measurement of the QCD inclusive jet cross section no deviations
with respect to the NLO theory prediction is observed, even up to the highest jet $p_{T}$ ever
probed of about 610 GeV. However, no explicit limit on $M_{q^{*}}$ is derived from this
measurement yet.

%
%%%
%
\section{Extended Gauge Symmetry}\label{sec:extd_gauge_sym}
\par
If there exists a grand unification of the strong and the electroweak interactions at a high energy
scale, then the breakdown of the corresponding gauge group (i.e. $SU(5)$, $SO(10)$, $E_{6}$,...)
downto the SM gauge group occurs through a cascade of symmetry breakings where extra
SU(2) and U(1) factors may appear. Such extra gauge group factors predict the existence
of new (and heavy since not observed yet) W and Z bosons that we respectively denote W' and Z'.
\par
\begin{figure}[h]
\begin{center}
\psfig{figure=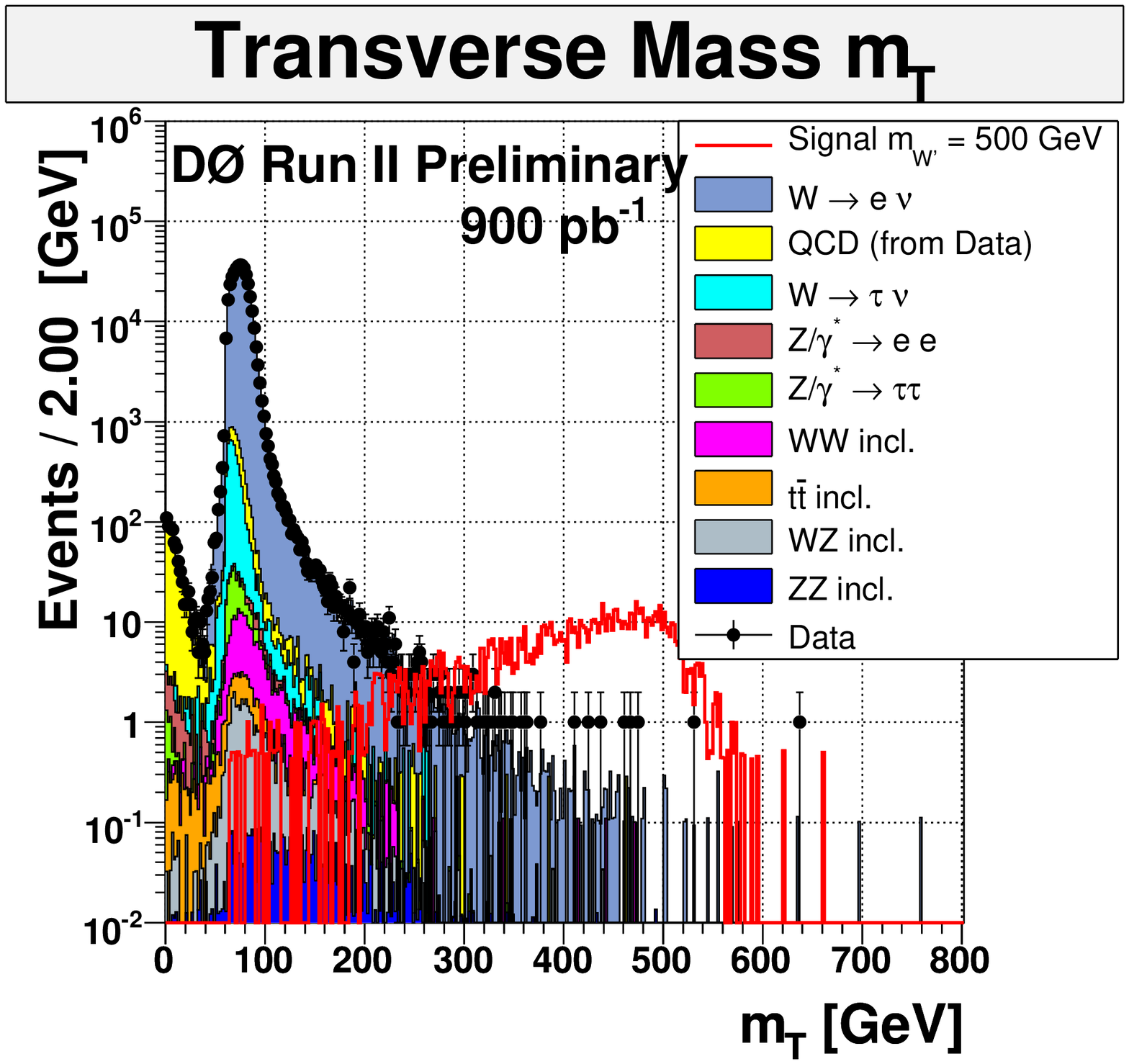,height=4cm}
\hspace{0.5cm}
\psfig{figure=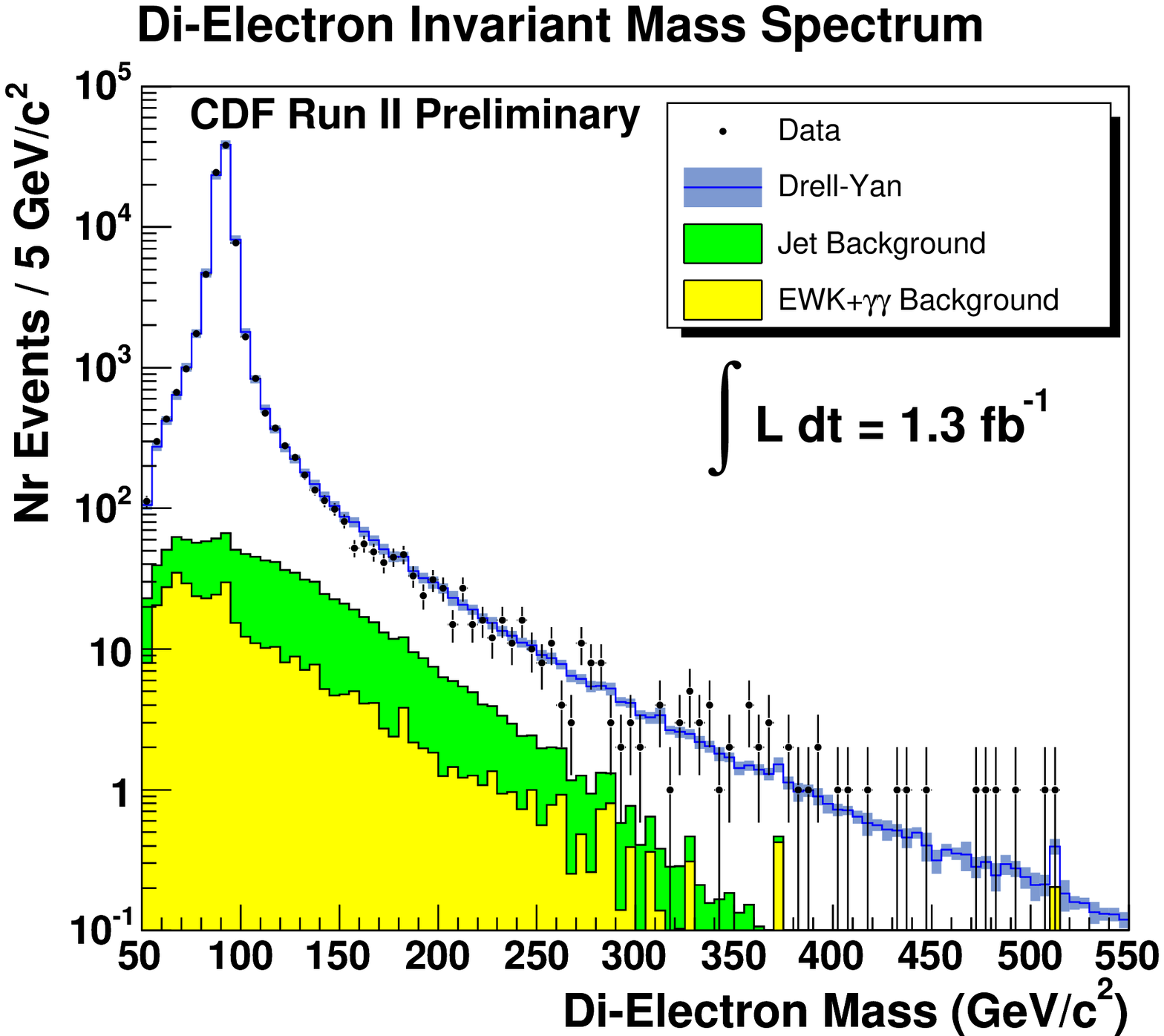,height=4cm}
\caption{Distributions of the transverse mass (left) and of the invariant mass (right) in searches
for W' by D0 and for Z' by CDF respectively.\label{fig:GUT}}
\end{center}
\end{figure}

We present a D0 search for the $W^{'\pm}\to e^{\pm}\nu$ process in 0.9 $fb^{-1}$ of data. The events
are selected if they contain a hard and isolated electron and a significant
$\rlap{\kern0.25em/}E_{T}$. The transverse mass $M_{T}(e^{\pm},\rlap{\kern0.25em/}E_{T})$ distribution
displayed on the left hand part of figure \ref{fig:GUT} is scrutinized especially above
150 GeV. No data excess is found on top of the SM background tail. Therefore a $W^{'\pm}$ with a mass
below 965 GeV is excluded. 
\par
We present a CDF search for the $Z^{'}\to e^{+}e^{-}$ process using a
data sample of 1.29 $fb^{-1}$. The analysis selects events with two hard and isolated electrons
with at least one in the central part of the calorimeter and with a matching track. Here the signal
region is defined as the tail (above 150 GeV) of the dielectron invariant mass shown on
the left hand side of figure \ref{fig:GUT}. The data are in good
agreement with the SM background causing a $Z^{'}$ with a mass lower than 923 GeV and SM-like
couplings to be excluded. 

%
%%%
%
\section{Extended Number of Space Dimensions}\label{sec:extd_space_dim}
\par
The lack of precision measurements of gravity in the sub-millimeter domain leaves some
room for possible departures from the Newton's law in this distance range. Such exotic behaviours are 
obviously predicted in the assumption that space has more than three dimensions because of the
Gauss law. This explains the apparent weakness of the gravitational interaction with respect to the other
fundamental interactions.
\par
The Randall-Sundrum (RS) model \cite{RS} postulates the existence of a fith dimension separating two branes. 
The SM fields are localized on one brane. Gravity lives on the second brane where it isn't weak, but can propagate 
along the fith dimension that has a warped metric. Kaluza-Klein excitations of the gravitons appear as narrow resonances.
The relevant parameters are the mass of the first graviton excitation and 
the $\frac{\kappa\sqrt{8\pi}}{M_{Pl}}$ coupling.
\par\noindent
The Arkani-Hamed, Dimopoulos and Dvali (ADD) model \cite{ADD} also localizes the SM fields on one 
brane and allows gravity to propagate within a bulk possibly made of up to $N=7$ large extra dimensions.
Here Kaluza-Klein excitations of the gravitons cannot be resolved. And the parameters are the number
of extra dimensions N and the effective Planck scale $M_{D}$ (i.e. the Planck scale in 4+N dimensions).
\par
\begin{figure}[h]
\begin{center}
\psfig{figure=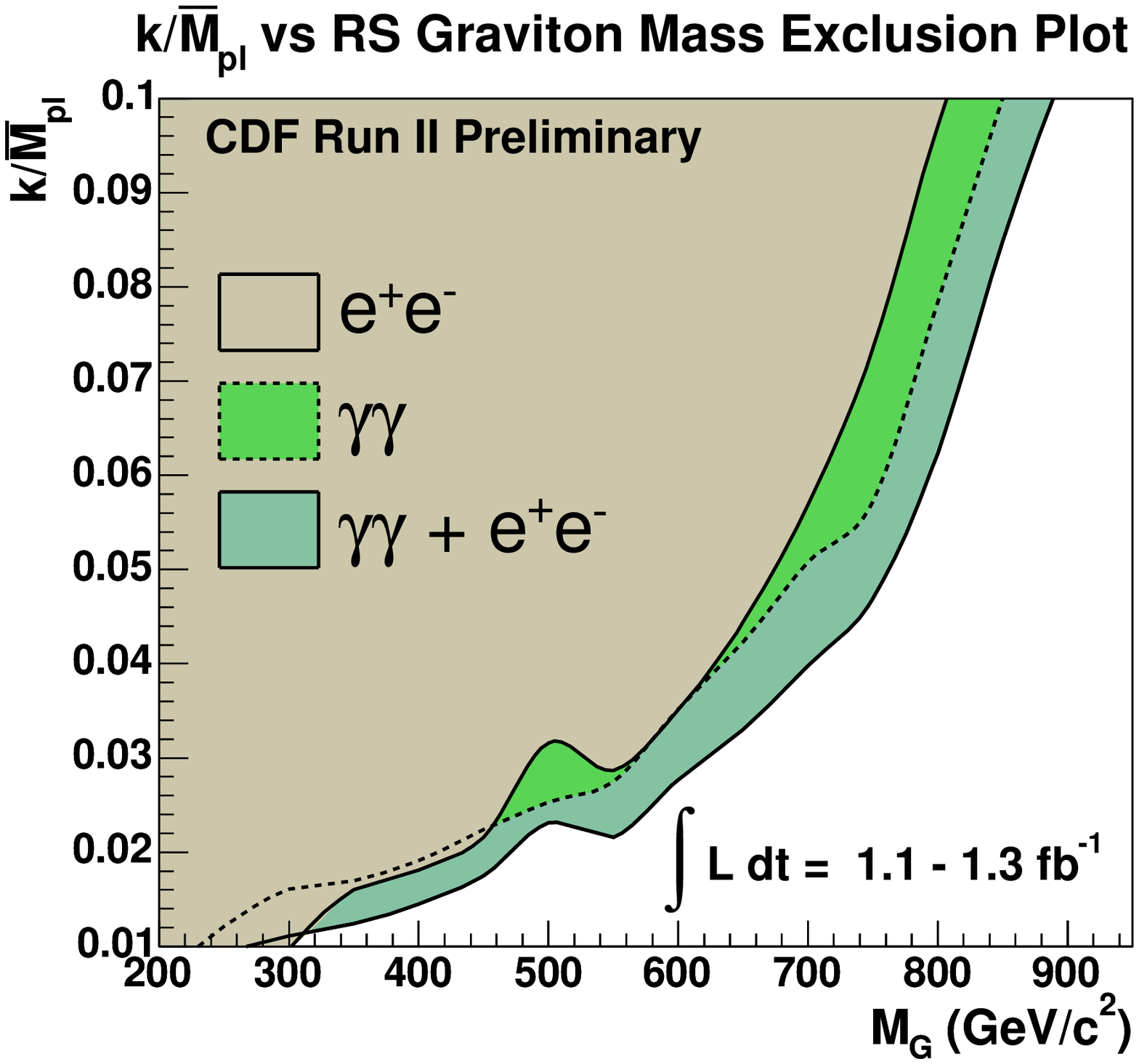,height=4cm}
\hspace{0.5cm}
\psfig{figure=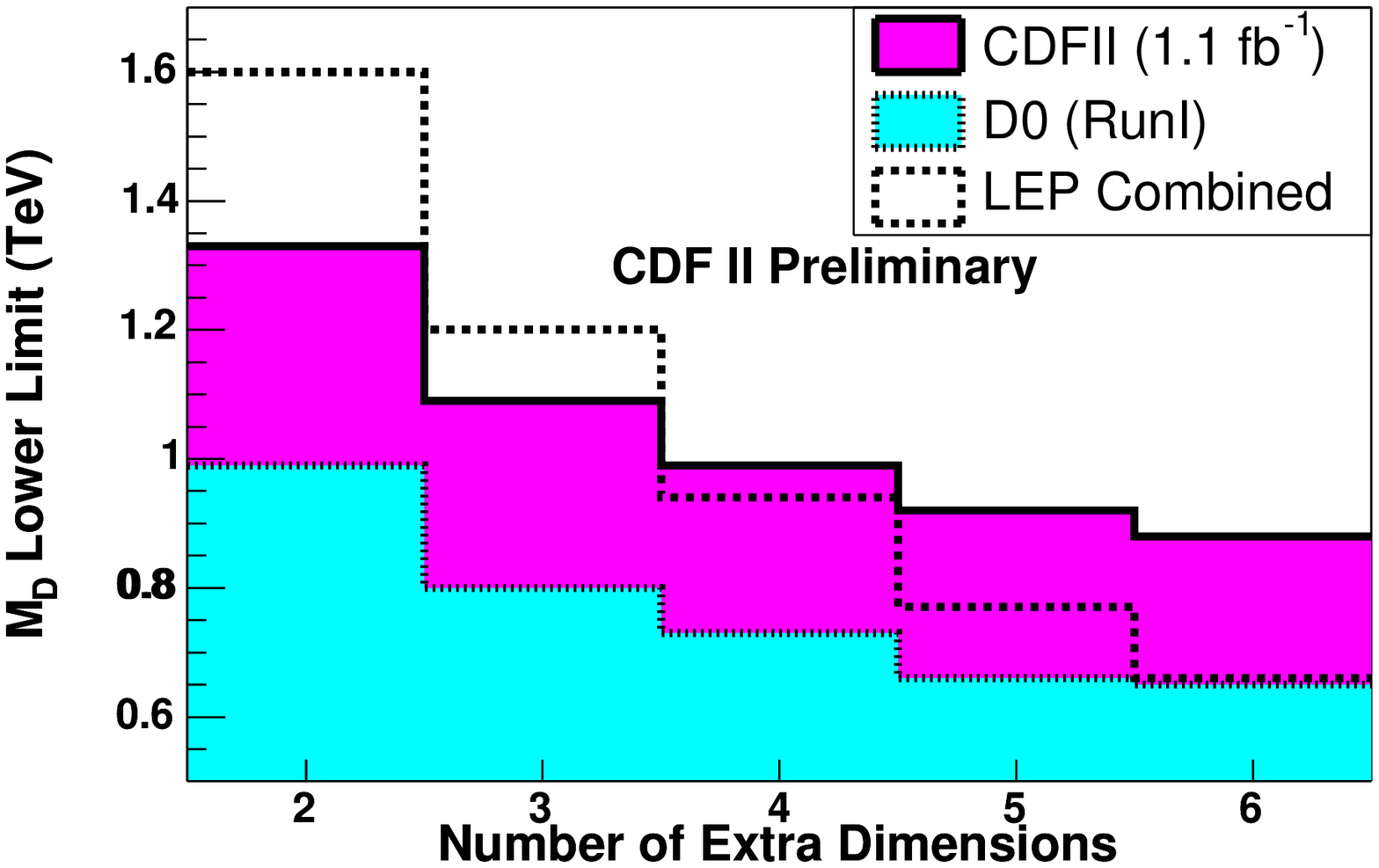,height=4cm}
\caption{Exclusion plot from the RS CDF search combining the $G\to e^{+}e^{-}$ and $G\to\gamma\gamma$
channels (left). Limits on large extra dimensions from the CDF monojet search (right).\label{fig:ED}}
\end{center}
\end{figure}

CDF recycled its $Z^{'}\to e^{+}e^{-}$ search into searches for $G\to e^{+}e^{-}$ and $G\to\gamma\gamma$.
The combination of the two latest yields the exclusion plane displayed at the left hand side of figure
\ref{fig:ED}.
\par\noindent
We also report on a CDF search in a $1.1 fb^{-1}$ dataset for the $q\bar q\to G+g$ contact interaction process.
This process leads to a monojet topology. Hence the analysis requires events with a very hard jet 
contained in the central
part of the calorimeter and confirmed by tracks. In order to allow for a gluon ISR or FSR a second soft
jet is accepted. The discriminating variable is the $\rlap{\kern0.25em/}E_{T}$ which is compatible with the
expected background. Consequently exclusion limits are derived as a function of the number of extra dimensions
and the effective Planck scale as shown at the right hand side of figure \ref{fig:ED}.
 
%
%%%
%
\section{Conclusion}
\par
Many searches covering very different topologies have been studied at the Tevatron Run II by the D0 and CDF
collaborations. Despite the recent updates of some of these analyses to the full Run IIA datasets of about $1 fb^{-1}$, no
hints of exotic extensions of the SM have been found and more stringent exclusion limits have been derived.
%
%%%
%

\end{document}